\documentclass[a4paper,11pt]{article}
\usepackage{jcappub}
\usepackage[T1]{fontenc}
\usepackage{caption}
\usepackage{subcaption}
\usepackage{amsmath, amssymb, graphics}
\usepackage{mathtools}
\usepackage{tablefootnote}
\usepackage{soul}
\usepackage{ulem}
\usepackage{color}
\usepackage{float}
\usepackage{makecell}
\usepackage{booktabs}
\usepackage[usenames,dvipsnames,svgnames,table]{xcolor}
\usepackage{indentfirst}
\usepackage{leftidx}
\usepackage{multirow}

\definecolor{LightCyan}{rgb}{0.88,1,1}
\newcolumntype{a}{>{\columncolor{LightCyan}}c}

\usepackage{xspace}

\usepackage{graphicx}
\usepackage{dcolumn}
\usepackage{bm}
\usepackage{subcaption}
\usepackage{amsmath}
\usepackage{amssymb}
\usepackage{mathtools}
\usepackage{tablefootnote}
\usepackage{soul}
\usepackage{ulem}
\usepackage{color}
\usepackage{float}
\usepackage{makecell}
\usepackage[usenames,dvipsnames,svgnames,table]{xcolor}
\usepackage{indentfirst}
\usepackage{leftidx}
\usepackage{slashed}

\title{Why is zero spatial curvature special?}

\author[1,2]{Raul Jimenez,}%
\author[3,1,4]{Ali Rida Khalife,}
\author[5]{Daniel F.~Litim,}
\author[6,7,8,9]{Sabino Matarrese,}
\author[10,11,12]{Benjamin D.~Wandelt}

\affiliation[1]{ICC, University of Barcelona, Marti  i Franques, 1, E-08028 Barcelona, Spain.}
\affiliation[2]{ICREA, Pg. Lluis Companys 23, Barcelona, E-08010, Spain.}
\affiliation[3]{Sorbonne Universit\'e, CNRS, UMR 7095, Institut d'Astrophysique de Paris, 98 bis bd Arago, 75014 Paris, France.}
\affiliation[4]{Dept. de  Fisica Cuantica y Astrofisica, University of Barcelona, Marti  i Franques 1, E-08028 Barcelona, Spain.}
\affiliation[5]{Department of Physics and Astronomy, University of Sussex, Brighton, BN1 9QH, U.K.}
\affiliation[6]{Dipartimento di Fisica e Astronomia Galileo Galilei, Universit\`a di Padova, 35131 Padova, Italy}
\affiliation[7]{INFN, Sezione di Padova, via F  Marzolo 8, I-35131 Padova, Italy.}
\affiliation[8]{INAF - Osservatorio Astronomico di Padova, vicolo dell Osservatorio 5, I-35122 Padova, Italy.}
\affiliation[9]{Gran Sasso Science Institute, viale F. Crispi 7, I-67100 L'Aquila, Italy.}
\affiliation[10]{Sorbonne Universit\'e, CNRS, UMR 7095, Institut d'Astrophysique de Paris, 98 bis bd Arago, 75014 Paris, France.}
\affiliation[11]{Sorbonne Universit\'e, Institut Lagrange de Paris (ILP), 98 bis bd Arago, 75014 Paris, France.}
\affiliation[12]{Center for Computational Astrophysics, Flatiron Institute, 162 5th Avenue, 10010, New York, NY, USA.
}%

\emailAdd{raul.jimenez@icc.ub.edu}
\emailAdd{ridakhal@iap.fr}
\emailAdd{d.litim@sussex.ac.uk}
\emailAdd{sabino.matarrese@pd.infn.it} 
\emailAdd{bwandelt@iap.fr}

\abstract{
Evidence for {\it almost} spatial flatness of the Universe has been provided from several observational probes, including the Cosmic Microwave Background (CMB) and Baryon Acoustic Oscillations (BAO) from galaxy clustering data. However, other than inflation, and in this case only in the limit of infinite time, there is no strong \textit{a priori} motivation for a spatially flat Universe. Using the renormalization group (RG) technique in curved spacetime, we present in this work a theoretical motivation for spatial flatness. Starting from a general spacetime, the first step of the RG, coarse-graining, gives a Friedmann-Lema\^itre-Robertson-Walker (FLRW) metric with a set of parameters. Then, we study the rescaling properties of the curvature parameter, and find that  zero spatial curvature of the FLRW metric is singled out as the unique scale-free, non-singular background for cosmological perturbations. }

\begin{document}
\maketitle

\section{Introduction \label{sec:intro}}

It has been known for a long time that the FLRW metric\footnote{We are using units in which the speed of light $c$ and the reduced Planck constant $\hbar$ are 1.},
\begin{equation}
    ds^2=-dt^2+a^2(t)\bigg(\frac{dr^2}{1-kr^2}+r^2d\theta^2+r^2\sin^2\theta d\phi^2\bigg),
    \label{Eq:FLRW_Metric}
\end{equation}
is the one that maximizes the symmetries of space~\cite{Landau:1975pou,Weinberg:1972kfs}; FLRW maximizes the symmetries of space, but only de-Sitter and Minkowski maximize symmetries of space-time. In the above equation, $ds^2$ is the line element, $t$ is cosmic time, $a(t)$ is the time dependent scale factor, $k$ is the curvature parameter and $\{r,\theta,\phi\}$ are the spatial spherical coordinates. Moreover, from the uniqueness theorem of maximally symmetric spaces, there is no other metric with further symmetries (see part four in chapter 13 of~\cite{Weinberg:1972kfs}). It is tempting to think this could be the reason why we live in a FLRW Universe. However, a further question arises: is there anything special about spatial flatness, i.e. $k=0$?

Observationally, analysis from the CMB~\cite{Planck:2018vyg,ACT:2020gnv,SPT-3G:2021wgf}, BAO~\cite{Efstathiou:2020wem,SDSS-DR7,BOSS:2016wmc} and the full shape of the large scale structure power spectrum~\cite{shapefit} have concluded that we live in a Universe with $|\Omega_{k0}|\lesssim0.001$, where $\Omega_{k0}=-kc^2/H_0^2$ is the curvature density parameter today, $H_0$ being the Hubble parameter today. Other interpretations of such data read it as evidence for a closed universe, with $\Omega_{k0}<0$~\cite{DiValentino:2019qzk,DiValentino:2020hov,Handley:2019tkm,Park:2017xbl}. Intriguingly, the majority of observational analyses in the literature  assume flatness  by setting $\Omega_{k0}=0$ from the outset.
Is there any compelling argument for the flat case to be singled out \textit{a priori} in this way?

So far, inflation is considered as the primary theoretical motivation for spatial flatness~\cite{Inflation1,Inflation3,Inflation4}. By describing a very early phase of accelerated expansion, most inflationary models point towards $k \approx 0$ today being an attractor solution for the metric of the Universe. In these models, deviations from flatness are at the level of one part in $10^{5}$ due to initial curvature fluctuations~\cite{Inflation5}, which could be measured at that level with future observations~\cite{racca}. 
An inflationary phase is the most prominent explanation for having a Universe consistent with spatial flatness today\cite{Inflation6}. However, it would be fruitful to have an additional theoretical motivation in favor of the specialty of a spatially flat universe (for alternative explanations of why the Universe is flat see e.g.~\cite{Turok}).

Stated in a different way, 
the curvature density parameter $\Omega_k \equiv -kc^2/a^2H^2$
can a priori take {\it any} negative value in the closed Universe case, {\it any} positive value in the open Universe case, but only 0 for the $k=0$ case.
This means that the flat universe case corresponds to a {\it set of measure zero} among all possible spatial geometries. 
Therefore, an exactly spatially flat Universe has zero probability to be selected among all possible FRLW models (a point that was recently argued in~\cite{2022arXiv220706547A}). 
This raises the question whether $\Lambda$CDM with $\Omega_k =0$ should be considered at all. It would seem that this requires theoretical arguments that select  $\Omega_k =0$ as an a priori special, maximally symmetric background for cosmological perturbations.

To that end, one might start by noting a crucial mathematical property of the spatial part of the FLRW metric, which is its conformal flatness (see~\cite{RoyMaha} and references therein), i.e. its Weyl tensor vanishes. This means that the spatial part of the FLRW metric can be transformed into the Minkowski one by an appropriate combination of coordinate and Weyl transformations~\footnote{It should be noted that these transformations are not technically diffeomorphisms, in the sense that the global properties of Minkowski spacetime are not recovered.}. However, while the cases $k>0$ and $k<0$ involve relatively complicated transformations to achieve that, as we will show below (see also section 111 of~\cite{Landau:1975pou}), the case $k=0$ becomes flat through a simple Weyl rescaling of the metric by $a^2(\eta)$, where $\eta$ is the conformal time. This simplicity could be hinting to a fundamental property of the case $k=0$ that makes it more favorable than the others. On the other hand, when studying the properties of a system under rescaling of the coordinates, the first thing that could come to mind is the renormalization group (RG) technique~\cite{Wislon1,Wislon2,Kadanoff:1966wm}. Although this technique is traditionally used to study critical phenomena in condensed matter physics~\cite{RG_and_Stat_Mech}, nothing prevents its use in the context of curved spacetime, as has been done before~\cite{Litim:2007pbw,Litim:2011cp,RG1,RG2}.

It is therefore the purpose of this work to apply the RG technique to a general spacetime, and show that $\Omega_k =0$ corresponds to a stable critical fixed-point of the RG flow; and indeed the only fixed point that is suitable as a background for cosmological perturbations. This provides  model-independent theoretical  motivation for the case of a flat FLRW universe.  

We start the discussion with a brief review of the RG, along with its steps (coarse-graining, rescaling and renormalization) and terminology(fixed-points, critical fixed-points, correlation length...) in section~\ref{sec:RG_Rev}. Then, we apply the coarse-graining step of the RG to a general spacetime in section~\ref{sec:Coarse_Graining} to get an FLRW-like metric. In section~\ref{sec:Res_Ren}, we recognise that $k$ has mass dimension 2 and  take inspiration from the RG to find fixed points under rescaling. This leads to examining a discrete number of fixed points. We then find that amongst these, the only non-singular solution that remains and that is suitable as a background for cosmological perturbations is $\Omega_k =0$. We end the discussion with a conclusion and summary in section~\ref{sec:Summ}. We include in appendix~\ref{sec:Appendix-A} some technical details.

\section{A brief review of RG \label{sec:RG_Rev}}

The main realm of application of the RG technique is in Statistical mechanics and Condensed Matter Physics, particularly to describe phase transitions~\cite{RG_and_Stat_Mech}. These abrupt transformations occur when the parameters of a certain system take particular values, known as the critical point of the system\footnote{It corresponds to a point in the space of  parameters, or equivalently the space of  Lagrangians or Hamiltonians.}, with a basic example being the ordered-disordered transition of the Ising spin model~\cite{Ising,Landau_Stat_Mech1}. Prior to the introduction of the RG method, the treatment of phase transitions was made using the mean field approximation (MFA), in which one considers small fluctuations around the average spin of the Ising model, for instance. However, although the MFA worked well in spatial dimension $d\geq4$, it broke down for small dimensions, where one could not even see a phase transition for $d=1$~\cite{tauber_2014,Tong}. This motivated the community to check for alternatives to the MFA, which brings us to the RG. 

We will briefly review the RG formalism and direct the interested reader to more detailed descriptions in~\cite{Simon,Tong,RG_and_Stat_Mech,Rosten2010vm}. The basic notion upon which the RG is based  is that the physical description of a system changes with the scale at which it is observed.
Therefore, one can start by studying the system at short distances, and the RG provides a step by step transition to larger distances~\cite{Rosten2010vm}. To understand how it works and its main outcome, let us start with a simple example.

Suppose we are interested in studying the transition to the ferromagnetic phase of a metal, which surely has a microscopic treatment. However, that would be too complicated if we are merely concerned with this phase transition due to thermal fluctuations. One can therefore define the average collection of spins about a point $\mathbf{x}$ as the magnetization $\mathbf{m}(\mathbf{x})$. In the presence of an external magnetic field $\mathbf{h}$, one can write a general\footnote{This coarse-grained Hamiltonian is based on the assumptions of locality, translational and rotational invariance.} Hamiltonian $H[\mathbf{m(x)}]$ as the Landau-Ginzburg one~\cite{Ginzburg:1950sr,Tong}:
\begin{align}
    \beta H[\mathbf{m(x)}]=\int d\mathbf{x}&\bigg[\frac{t}{2}\mathbf{m}^2+u\mathbf{m}^4+...+\frac{K}{2}(\nabla\mathbf{m})^2\nonumber\\
    &+\frac{L}{2}(\nabla^2\mathbf{m})^2+\frac{N}{2}\mathbf{m}^2(\nabla\mathbf{m})^2+...\nonumber\\
    &-\mathbf{h}.\mathbf{m}\bigg]
    \label{Eq:LG Hamiltonian}
\end{align}
where $\beta=1/(k_BT)$, with $k_B$ the Boltzmann constant and $T$ the temperature. Plus, the coefficients $t,u,K...$ are phenomenological parameters for their respective terms, and they define the axes of the theory space. Each point in this space corresponds to a different Hamiltonian, and thus a different theory. From here, one can then apply the recipe of the RG as follows: 
\begin{enumerate}
    \item \textit{Coarse-Graining:} As mentioned previously, the study of phase transitions is more relevant on distances larger than the microscopic scale $a$, i.e. the lattice spacing. Therefore, we first must decrease the resolution by averaging out fluctuations in $\mathbf{m}(\mathbf{x})$ that are smaller than $ba$, 
    \begin{equation}
        \bar{\mathbf{m}}(\mathbf{x})=\frac{1}{(ba)^d}\int_{C} d\mathbf{y}\ \mathbf{m}(\mathbf{y})
    \end{equation}
    where $b>1$ is the rescaling parameter, $d$ is the number of spatial dimensions and the integral is over a cell $C$ of size $(ba)^d$ centered on $\mathbf{x}$. Note that this is equivalent to integrating out large momentum modes.
    \item \textit{Rescaling:} By changing the resolution, one now has a ``grainier'' picture than the original one, and therefore one cannot directly link the two. In order to restore that, one simply rescales the coordinates $\mathbf{x}\rightarrow b\mathbf{x}$.
    \item Finally, due to the above two steps, the transformed theory will end up with some additional coefficient in front of its kinetic term, call it $\zeta^2$. This could be interpreted as a change in the contrast between the two theories. In order to return it to its canonical form, one can simply rescale the magnetization $\mathbf{m}(\mathbf{x})\rightarrow\zeta\bar{\mathbf{m}}(b\mathbf{x)}$.
\end{enumerate}
This operation can be compactly stated as $\mathbf{S}\rightarrow \mathbf{S'}=\mathbf{R}_b\mathbf{S}$, where $\mathbf{R}_b$ is the RG transformation from one point in theory space($\mathbf{S}$) to another($\mathbf{S'}$) by a scaling parameter $b$. Note that symmetries are automatically preserved under $\mathbf{R}_b$\footnote{That doesn't mean that there's no symmetry enhancement at the critical point, as we will see shortly.}, and so the new $H_{\zeta}$ has the same form as the old $H$ but in terms of the renormalized variables. Moreover, Hamiltonians that are statistically self-similar correspond to ``fixed points'' in parameter space. They are invariant under the RG: if $S^{*}$ is a fixed point then $S^{*} = \mathbf{R}_{b} S^{*}$. These points are crucial because they are identified with the phase transition points; this is where the link between RG and thermodynamics appears. More specifically, since under RG the correlation length\footnote{The correlation length is the distance over which field fluctuations are correlated. It thus defines the length scale up to which we have self-similarity.} $\xi\rightarrow b\xi$, this means that at the fixed point, $\xi$ can be either zero or infinite. The latter corresponds to critical phase transitions.

A seminal example for critical points is given by the $3d$ Ising model \cite{Wilson:1971dc}, where the parameters of the effective Hamiltonian \eqref{Eq:LG Hamiltonian} may take fixed points under the RG transformation. Concretely, in terms of the effective mass term $\sim t$, the model is known to display the free Gaussian fixed point $(t_*=0$), the infinite Gaussian fixed point ($1/t_*=0$), and the seminal Wilson-Fisher fixed point ($t_*=$ finite) characterising a second order phase transition with a diverging correlation length  \cite{Nicoll:1974zza}.

Having identified the fixed points of the theory space, one can then study its stability by linearizing deviations about it:
\begin{equation}
    S_{i}^{*} + \delta S_{i}^{'} = S_{i}^{*} + (\mathbf{R}_b)_{i j} \delta S^{j}
\end{equation}
where $(\mathbf{R}_b)_{i j} = \frac{\partial S_{i}^{'}}{\partial S^{j}}$. Since $\mathbf{R}_b$ is a transformation, one can find its eigenvalues and eigenvectors $\lambda_i(b) = b^{y_i}$ and $\mathbf{\Theta}_i$, respectively, where $y_i$ is called the anomalous dimension of operator $\mathbf{\Theta}_i$. For $y_i >0,<0$ or 0, the operator $\mathbf{\Theta}_i$ is said to be a relevant (RO), irrelevant (IO) or marginal operator (MO), respectively. If we imagine moving away from a fixed point in theory space according to a flow dictated by these operators, the IOs are those that will take us back to the original fixed point. That is why IOs are crucial to identify stable fixed points which define a subspace in theory space called the ``basin of attraction".

In many areas of physics, the RG approach has unveiled  fundamental properties of the laws of nature. In this context, it is easy to understand what a critical component of a phase transition is as the  critical fixed point of the renormalization group. This is universal no matter what the nature of the material undergoing the phase transition is. In particular, one can imagine applying this formalism to other, more abstract physical entities, such as spacetime and its metric, which we will now discuss.

\section{RG in Curved Spacetime\label{sec:RG_Curv}}

After the brief review of the RG technique, we will now apply it to the metric of spacetime. Our theory space will be that of the metric components, in analogy to the Hamiltonian of Sec.~\ref{sec:RG_Rev}, and the metric. 
As a starting point, we use completely general metrics in four space-time dimensions without assuming any form of symmetry. We further assume that General Relativity (GR) is the underlying theory of gravity. We then argue that FLRW metrics \eqref{Eq:FLRW_Metric} with vanishing spatial curvature are distinguished, much like critical points of  the RG.

\subsection{Coarse Graining\label{sec:Coarse_Graining}}

In the present context, coarse-graining amounts to averaging over small distances, and then to look at how this modifies the large distance behaviour. To do this for curved spacetime metrics, we adopt the machinery presented in~\cite{Buchert:2000,Buchert:2001,Sabino:2005}. 
The latter assumes spacetime to be filled with matter in the form of dust or a perfect fluid, which is a good description for all  known components of our universe -- baryons, dark matter, dark energy, and radiation. 
We will not go through the mathematical details here, but rather we focus on the main principles and results needed for our purpose.

We then consider an irrotational cosmic fluid and foliate spacetime into hypersurfaces orthogonal to its 4-velocity.\footnote{Note that the irrotational assumption can be relaxed and the final results do not depend on the particular foliation, see Sec.~2.1 of~\cite{Buchert:2001}.} This allows us to write 
the line element as
\begin{equation}\label{eq:general}
    ds^2=-N^2dt^2+g_{ij}dX^idX^j,
\end{equation}
where $N$ is the lapse function, $g_{ij}$ are the spatial components of the metric and $X^i$ are called Lagrangian coordinates. Next, we consider a compact and simply-connected domain\footnote{A compact domain is one that is closed and bounded,i.e. it has no holes or missing limiting points. For e.g. the set $[0,1]$ is compact while $[0,1]-\{0.2\}$ is not since it has a hole (it's missing the 0.2). A simply connected domain is one without any holes in it~\cite{Real_Analysis}.} $\mathcal{D}$ contained within spatial hypersurfaces of constant time to define the spatial average  of a quantity $\Psi$ as 
\begin{equation}\label{eq:CG}
    \langle\Psi\rangle_{\mathcal{D}}\equiv\frac{1}{V_{\mathcal{D}}}\int_{\mathcal{D}}\Psi \,\sqrt{\det g_{ij}}\, d^3X,
\end{equation}
where the domain's volume  
$V_{\mathcal{D}}$ is given by 
\begin{equation}
    V_{\mathcal{D}}(t)\equiv\int_{\mathcal{D}}\, \sqrt{\det g_{ij}}\,d^3X\,.
\end{equation}

Further, using the definition of $V_{\mathcal{D}}(t)$, one can then extract a dimensionless scale factor,
\begin{equation}
    a_{\mathcal{D}}(t)\equiv\bigg(\frac{V_{\mathcal{D}}(t)}{V_{\mathcal{D}_0}(t)}\bigg)^{1/3},
\end{equation}
normalized to the volume of the initial domain $V_{\mathcal{D}_0}$. With this scale factor, we are simply concerned with the effective dynamics over the domain. Intriguingly, continuing the process of averaging the equations of motion over increasingly larger volumes
{\it invariably} leads to
modified Friedmann equations (see eqs.(13a,b) in~\cite{Buchert:2001}). 
These include a curvature term $k$, in addition to kinematical and dynamical backreaction terms $Q$ and $P$, respectively 
\cite{Buchert:2001,Paranjape:2006ww}. We conclude that maximally symmetric FLRW-like metrics \eqref{Eq:FLRW_Metric}   arise  from averaging  general spacetimes such as \eqref{eq:general}, with the curvature parameter $k$ and the backreaction parameters $Q$ and $P$ becoming part of the theory space of parameters.

\subsection{Scaling analysis\label{sec:Res_Ren}}

We are now in a position to perform a scaling analysis of 
\eqref{Eq:FLRW_Metric} in the sense of the RG. 
On dimensional grounds, starting with \eqref{Eq:FLRW_Metric}, we observe that the product $k\,r^2$ must be dimensionless. 
Given that the canonical mass dimension of a length is $[r]=-1$, 
it follows that the canonical mass dimension of curvature is $[k]=2$, much like the mass term $t$ in the model Hamiltonian \eqref{Eq:LG Hamiltonian}. We also have that $[a]=0$, by definition of the metric. It then follows that the (classical) running of curvature with coarse-graining scale takes the form
\begin{equation}\label{eq:curvature}
    \ell \partial_{\ell} \,\hat{k} = 2 \hat{k}\,,
\end{equation}
with $\hat k = k \ell^2$ the dimensionless version of the curvature parameter. Hence, we observe two scale-free classical fixed points, 
the Gaussian fixed point  $(\hat{k}_* =0)$, and the infinite Gaussian fixed point $(1/\hat{k}_* =0)$ \cite{Nicoll:1974zza}. While the former is self-evident, the latter becomes visible using the mapping $\hat{k} \rightarrow 1/\hat{k}$. These classical fixed points can be viewed as the counterparts of the Gaussian and infinite Gaussian fixed point of the mass term in Ising-type models indicated earlier, and characterise the asymptotic limits where spatial curvature  is either absent, or fully dominant.  

This also reflects the well-known results that, firstly, in general relativity, the flat solution is unstable under IR perturbations (also known as the ``flatness'' problem), and, secondly, that  the curvature of the universe is of no relevance when we run particle detectors (i.e. in the UV limit).

As an aside, we note that the analogue of a Wilson-Fisher fixed point may arise if spatial curvature itself, by virtue of the successive coarse graining steps \eqref{eq:CG}, depends on the coarse graining  scale, $k=k(\ell)$. The latter would generate an additional term $\sim \ell^3\partial_\ell k(\ell)$ on the right-hand side of \eqref{eq:curvature}, which may offer non-trivial fixed points. We defer further exploration of such non-trivial fixed points to future work.

Next, we relate our discussion to the conformal flatness of FLRW space-times. Applying the coordinate transformation 
\begin{equation}
    r \rightarrow \frac{r}{1+k \, r^2 /4}
    \label{Eq:transformation}
\end{equation}
to \eqref{Eq:FLRW_Metric}  entails that the spatial metric (or line element $ds_3^2$) can be written as 
\begin{equation}
    d s^{2}_{3} = a^2 \left (1+\frac{k r^2}{4} \right )^{-2} \left [ dr^2 + r^2 d\theta^2 + r^2 \sin^2 \theta d\phi^2 \right ]
    \label{Eq:new_metric}
\end{equation}
(see the Appendix~ for derivation).
The conformal flatness of the spatial part of the FLRW is now manifest as~\eqref{Eq:new_metric} coincides with the flat space metric, except for an overall $r$-dependent factor. 
Moreover, notice that a characteristic length scale 
\begin{equation}
    r_k\equiv\frac{2}{\sqrt{|k|}}
\end{equation}
automatically comes out of this transformation. It states  that space looks effectively flat for distances $r\ll r_k$, while it appears significantly curved for   distances $r\gg r_k$. 
We also note that $r_k\to\infty$ from the outset if and only if spatial curvature vanishes $k=0$. 
For the purpose of this study,  we are mainly interested in the range $r\leq \ell_D$, where $\ell_D$  is the scale up to which spatial sections of the initial (foliated) space-time have been spatially averaged.
It follows that for $k \neq 0$, 
FLRW space times will look spatially curved rather than flat  at asymptotic distances. However, the transition from flat to curved disappears in the limit $k \rightarrow 0$, in which $r_k\rightarrow\infty$, whence vanishing spatial curvature can be interpreted as a critical point. 

\subsection{A background for cosmological perturbations\label{sec:CosmologicalBackgrounds}}

Having shown that the mass dimension of $k$ naturally leads to a scale of the FLRW universe, the curvature radius $r_k$, we will now show that removing this scale by rescaling uniquely leads to a scale-free, symmetric spacetime suitable as a background for cosmological perturbations. In the following we will use scaling arguments inspired by the RG approach described above.

However, it should be noted that the application differs in the following way. Usually, the volume average is done to define a coarse grained field and the relevant equations for perturbations which lead to predictions for coarse-grained observables. In contrast, here we are attempting to define a metric that can serve as a background for cosmological perturbation theory. Buchert's approach~\cite{Buchert:2000,Buchert:2001} shows that using a volume average starting from a general space-time metric achieves an FLRW effective ``background" geometry.   

Let us now present our argument. The  case of vanishing spatial curvature $k=0$ is already scale-free, so we focus on $k\neq 0$. In either cases we have two options to remove the curvature scale by achieving $r_k\rightarrow0$ or $r\rightarrow\infty$. This leads to the following cases, as can be seen from~\eqref{Eq:new_metric}. For either $k>0$ or $k<0$, the limit of $r_k\rightarrow\infty$ recovers $\Omega_k =0$ except with non-trivial global topology. Moreover, for positive spatial curvature $k>0$, the limit $r_k\rightarrow0$ leads to a singular solution with zero volume of the spatial sections. On the other hand, negative spatial curvature $k<0$  corresponds to a FLRW universe with hyperbolic spatial sections of infinite volume. Further, the limit $r_k \rightarrow 0$ here is clearly singular and, as such, unsuitable as a background for cosmological perturbations\footnote{An hyperbolic spatial section  with infinite curvature evokes the notion of a fractal universe that is nowhere smooth, but we will leave a detailed mathematical exploration of this limit to future work.}. 
In consequence, the unique scale-free symmetric background is one that is locally indistinguishable from a flat FLRW universe with $\Omega_k =0$, i.e. one that is identical to the latter on any finite patch, however large. 

\section{Summary\label{sec:Summ}}

Observationally, it will be impossible to prove conclusively that spatial curvature of the universe vanishes exactly, since any constraint on $\Omega_{k0}=0$ is limited by the cosmic variance value of $10^{-5}$. Theoretically, however, one might wonder whether arguments can be found 
that single-out spatial flatness ($k=0$).

Here, we have demonstrated that amongst all maximally symmetric  and scale free background metrics that emerge from  renormalization group arguments, the $\Omega_k=0$ FLRW universe is the only one that is suitable as a background for cosmological perturbations.

Let us clearly state what we do \textit{not} claim. We do not propose a  mechanism that dynamically generates such background solutions based on physical assumptions; in this sense the claim in this paper should not be seen as proposing an alternative to models of the very early universe, such as inflation, that aim to motivate a nearly flat and perturbed FLRW background. We did not consider any effect of possible  fluctuations as we were concerned only  with pure background effects;  the effect of fluctuations on the background  has been done by one of us (SM).  In~\cite{Sabino:2005} they considered the RG to investigate how super and sub-horizon fluctuations behave. Their eq. (64) shows that super-horizon fluctuations are just curvature; this is well establsihed. For the sub-horizon perturbations the situation is more complicated, the recent findings from relativistic numerical simulations is that probably the effect is small but even the calculation in their section 3.2 shows that this effect will be small at best and in no case will affect the fixed points of the RG.

However, our result does indicate that it \textit{is} admissible to accord flat solutions an \textit{a priori} special status even though they are a set of measure zero amongst all homogeneous and isotropic solutions to Einstein's equations. This may include explicitly focusing on flat models when exploring cosmological parameter constraints.

\begin{acknowledgments}
 We thank Cesar Gomez and Roy Maartens for very useful feedback on drafts of this paper. Funding for this work was partially provided by project PGC2018-098866-B-I00  MCIN/AEI/10.13039/501100011033 y FEDER ``Una manera de hacer Europa'', and the ``Center of Excellence Maria de Maeztu 2020-2023'' award to the ICCUB (CEX2019-000918-M funded by MCIN/AEI/10.13039/501100011033). 
 This work was performed in part at Aspen Center for Physics, which is supported by National Science Foundation grant PHY-1607611, and was partially supported by a grant from the Simons Foundation.
ARK acknowledges receiving funding from the European Research Council (ERC) under the European Union's Horizon 2020 research and innovation program (grant agreement No 101001897). 
 DFL is supported by the Science and Technology Research Council (STFC) under the Consolidated Grant ST/T00102X/1.
 SM acknowledges financial support from the COSMOS network (www.cosmosnet.it) through the ASI (Italian Space Agency) Grants 2016-24-H.0, 2016-24-H.1-2018 and 2019-9-HH.0. BDW acknowledges support by the ANR BIG4 project, grant ANR-16-CE23-0002 of the French Agence Nationale de la Recherche; and the Labex ILP (reference ANR-10-LABX-63) part of the Idex SUPER, and received financial state aid managed by the Agence Nationale de la Recherche, as part of the programme Investissements d'avenir under the reference ANR-11- IDEX-0004-02. The Flatiron Institute is supported by the Simons Foundation. 
\end{acknowledgments}

\vspace*{0.5cm}

\appendix
\section{Derivation of transformation~\eqref{Eq:new_metric}\label{sec:Appendix-A}}
The standard form of FLRW metric (spatial part) in spherical coordinates reads
\begin{equation}
    ds^2=a^2\bigg(\frac{dr^2}{1-kr^2}+r^2d\theta^2+r^2\sin^2\theta d\phi^2\bigg).
    \label{Eq:Stdr_FLRW}
\end{equation}
Consider $r={r_1}/({1+\frac{kr_1^2}{4}})$, leading to
\begin{align}
    dr=\frac{1-\frac{kr_1^2}{4}}{\bigg(1+\frac{kr_1^2}{4}\bigg)^2}dr_1.
\end{align}
Plug into~\eqref{Eq:Stdr_FLRW}
\begin{align}
    ds^2=& a^2\bigg[\frac{1}{1-\frac{kr_1^2}{\bigg(1+\frac{kr_1^2}{4}\bigg)^2}}\times\frac{\bigg(1-\frac{kr_1^2}{4}\bigg)^2}{\bigg(1+\frac{kr_1^2}{4}\bigg)^4}dr_1^2+\frac{r_1^2}{\bigg(1+\frac{kr_1^2}{4}\bigg)^2}d\theta^2\nonumber\\
    &+\frac{r_1^2}{\bigg(1+\frac{kr_1^2}{4}\bigg)^2}\sin^2\theta d\phi^2\bigg]\nonumber\\
    &=a^2\bigg(1+\frac{kr_1^2}{4}\bigg)^{-2}\bigg[\frac{\bigg(1-\frac{kr_1^2}{4}\bigg)^2}{\bigg(1+\frac{kr_1^2}{4}\bigg)^2-kr_1^2}dr_1^2+...\bigg]\nonumber\\
    &=a^2\bigg(1+\frac{kr_1^2}{4}\bigg)^{-2}\bigg[dr_1^2+r_1^2d\theta^2+r_1^2\sin^2\theta d\phi^2\bigg]
\end{align}
which is~\eqref{Eq:new_metric} above.



\providecommand{\noopsort}[1]{}\providecommand{\singleletter}[1]{#1}%

\providecommand{\href}[2]{#2}\begingroup\raggedright\endgroup

\end{document}